\documentclass[12pt]{JHEP}

\usepackage{amsmath,epsfig}
\usepackage{amssymb,amsfonts}
\usepackage{latexsym}
\usepackage{longtable}

\relax

\def\be{\begin{equation}}
\def\ee{\end{equation}}
\def\bea{\begin{eqnarray}}
\def\eea{\end{eqnarray}}

\newcommand\fverb{\setbox\pippobox=\hbox\bgroup\verb}
\newcommand\fverbdo{\egroup\medskip\noindent%
                        \fbox{\unhbox\pippobox}\ }
\newcommand\fverbit{\egroup\item[\fbox{\unhbox\pippobox}]}

\newcommand{\la}{\lambda}

\newcommand{\bear}{\begin{eqnarray}}

\newcommand{\eear}{\end{eqnarray}}

\newbox\pippobox

\def\6{\partial}

\def\a{\alpha}

\def\sq
\def\a{\alpha}

\def\bx{{\bf x}}

\title{Ho\v{r}ava-Lifshitz Dark Energy}
\author{{\large Emmanuel N. Saridakis }
~\\
~\\
Department of Physics, University of Athens, GR-15771 Athens,
Greece {\tt msaridak@phys.uoa.gr}}

\preprint{}

\abstract{We formulate  Ho\v{r}ava-Lifshitz cosmology with an
additional scalar field that leads to an effective dark energy
sector. We find that, due to the inherited features from the
gravitational background,  Ho\v{r}ava-Lifshitz dark energy
naturally presents very interesting behaviors, possessing a
varying equation-of-state parameter, exhibiting phantom behavior
and allowing for a realization of the phantom divide crossing. In
addition, Ho\v{r}ava-Lifshitz dark energy guarantees for a bounce
at small scale factors and it may trigger the turnaround at large
scale factors, leading naturally to cyclic cosmology.}

\begin{document}

\section{Introduction}

Recently, a power-counting renormalizable, ultra-violet (UV)
complete theory of gravity was proposed by Ho\v{r}ava in
\cite{hor2,hor1,hor3,hor4}. Although presenting an infrared (IR)
fixed point, namely General Relativity, in the  UV the theory
possesses a fixed point with an anisotropic, Lifshitz scaling
between time and space of the form $\bx\to\ell~\bx$,
$t\to\ell^z~t$, where $\ell$, $z$, $\bx$ and $t$ are the scaling
factor, dynamical critical exponent, spatial coordinates and
temporal coordinate, respectively.

Due to the these novel features, there has been a large amount of
effort in examining and extending the properties of the theory
itself
\cite{Volovik:2009av,Cai:2009ar,Cai:2009dx,Orlando:2009en,Nishioka:2009iq,Konoplya:2009ig,Charmousis:2009tc,Li:2009bg,Nojiri:2009th,Kluson:2009rk,Visser:2009fg,Sotiriou:2009gy,
Sotiriou:2009bx,Germani:2009yt,Chen:2009bu,Chen:2009ka,Shu:2009gc,Bogdanos:2009uj,Afshordi:2009tt,Myung:2009ur,Alexandre:2009sy,
Blas:2009qj,Capasso:2009fh,Dutta:2009jn,Chen:2009vu,Kluson:2009xx,Dutta:2010jh}.
Additionally, application of Ho\v{r}ava-Lifshitz gravity as a
cosmological framework gives rise to Ho\v{r}ava-Lifshitz
cosmology, which proves to lead to interesting behaviors
\cite{Calcagni:2009ar,Kiritsis:2009sh}. In particular, one can
examine specific solution subclasses
\cite{Lu:2009em,Nastase:2009nk,Colgain:2009fe,Ghodsi:2009rv,Minamitsuji:2009ii,Ghodsi:2009zi,Wu:2009ah,Cho:2009fc,Boehmer:2009yz,Momeni:2009au,Setare:2009sw,Kiritsis:2009vz,Cai:2010ud},
the phase-space behavior
\cite{Carloni:2009jc,Leon:2009rc,Myung:2009if,Bakas:2009ku,Myung:2010qg},
the gravitational wave production
\cite{Mukohyama:2009zs,Takahashi:2009wc,Koh:2009cy,Park:2009gf,Park:2009hg,Myung:2009ug},
the perturbation spectrum
\cite{Mukohyama:2009gg,Piao:2009ax,Gao:2009bx,Chen:2009jr,Gao:2009ht,Cai:2009hc,Wang:2009yz,Kobayashi:2009hh,Wang:2009azb},
the matter bounce
\cite{Brandenberger:2009yt,Brandenberger:2009ic,Cai:2009in,Suyama:2009vy,Czuchry:2009hz,Gao:2009wn},
the black hole properties
\cite{Danielsson:2009gi,Cai:2009pe,Myung:2009dc,Kehagias:2009is,Mann:2009yx,Bertoldi:2009vn,Park:2009zra,Castillo:2009ci,BottaCantcheff:2009mp,Lee:2009rm,Ding:2009pq,Varghese:2009xm,Kiritsis:2009rx,Greenwald:2009kp,Lobo:2010hv},
the dark energy phenomenology
\cite{Park:2009zr,Appignani:2009dy,Setare:2009vm,Garattini:2009ns,Chaichian:2010yi},
the astrophysical phenomenology
\cite{Kim:2009dq,Harko:2009qr,Iorio:2009qx,Iorio:2009ek,Izumi:2009ry},
the thermodynamic properties
\cite{Wang:2009rw,Cai:2009qs,Cai:2009ph} etc.

In the present form of Ho\v{r}ava-Lifshitz cosmology, one combines
the aforementioned modified gravitational background with a scalar
field that reproduces (dark) matter. Doing so he obtains a
dark-matter universe, with the appearance of a cosmological
constant and an effective ``dark radiation'' term. Although these
terms are interesting cosmological artifacts of the novel features
of Ho\v{r}ava-Lifshitz gravitational background, they still
restrict the possible scenarios of Ho\v{r}ava-Lifshitz cosmology.

In the present work we are interested in formulating
Ho\v{r}ava-Lifshitz cosmology in a way that an effective dark
energy, with a varying equation-of-state parameter, will emerge.
Thus, we add a second scalar field, which dynamics will be
combined with the cosmological constant and dark radiation terms.
Although such a scalar field could put into question the
renormalizability of the theory, it is still interesting to
investigate what would be its effects on the cosmological behavior
of the universe. Indeed, it proves that the effective
Ho\v{r}ava-Lifshitz dark energy that arises, can have very
interesting cosmological implications at both early and late
times, such are to trigger a bounce and a turnabout. It's
equation-of-state parameter can present quintessence behavior, or
surprisingly enough, it can quite generally  give rise to phantom
behavior and to the crossing of the phantom divide. Although we
construct it under the detailed-balance condition, the basic
features of the model at hand are independent of that and are
expected to be present even if we relax this constraint. This
paper is organized as follows. In section \ref{model} we formulate
Ho\v{r}ava-Lifshitz cosmology with both dark matter and dark
energy fields. In section \ref{discuss} we examine the
cosmological behavior of the model and we discuss its
implications. Finally, section \ref{concl} is devoted to the
summary of the obtained results.

\section{Ho\v{r}ava-Lifshitz cosmology with dark matter and dark energy fields}
\label{model}

We begin with a brief review of Ho\v{r}ava-Lifshitz gravity.
 The dynamical variables are the lapse and shift
functions, $N$ and $N_i$ respectively, and the spatial metric
$g_{ij}$ (roman letters indicate spatial indices). In terms of
these fields the full metric is
\begin{eqnarray}
ds^2 = - N^2 dt^2 + g_{ij} (dx^i + N^i dt ) ( dx^j + N^j dt ) ,
\end{eqnarray} 
where indices are raised and lowered using $g_{ij}$. The scaling
transformation of the coordinates reads (z=3):
\begin{eqnarray}
 t \rightarrow l^3 t~~~{\rm and}\ \ x^i \rightarrow l x^i~.
\end{eqnarray}

Decomposing the gravitational action into a kinetic and a
potential part as $S_g = \int dt d^3x \sqrt{g} N ({\cal L}_K+{\cal
L}_V)$, and under the assumption of detailed balance \cite{hor3},
which apart form reducing the possible terms in the Lagrangian it
allows for a quantum inheritance principle \cite{hor2} (the $D+1$
dimensional theory acquires the renormalization properties of the
D-dimensional one),
 the full action of Ho\v{r}ava-Lifshitz gravity is given by
\begin{eqnarray}
&&S_g =  \int dt d^3x \sqrt{g} N \left\{ \frac{2}{\kappa^2}
(K_{ij}K^{ij} - \lambda K^2)
 - \frac{\kappa^2}{2 w^4} C_{ij}C^{ij}
 + \frac{\kappa^2 \mu}{2 w^2}
\frac{\epsilon^{ijk}}{\sqrt{g}} R_{il} \nabla_j R^l_k-
     \right. \nonumber \\
&&\left.\ \ \ \ \ \ \ \  \ \ \ \ \ \ \ \ \ \ \ \ \ \ \ \ \ -
\frac{\kappa^2 \mu^2}{8} R_{ij} R^{ij} + \frac{\kappa^2 \mu^2}{8(1
- 3 \lambda)} \left[ \frac{1 - 4 \lambda}{4} R^2 + \Lambda  R - 3
\Lambda ^2 \right] \right\},
\end{eqnarray}
where
\begin{eqnarray}
K_{ij} = \frac{1}{2N} \left( {\dot{g_{ij}}} - \nabla_i N_j -
\nabla_j N_i \right) \, ,
\end{eqnarray}
is the extrinsic curvature and
\begin{eqnarray} C^{ij} \, = \, \frac{\epsilon^{ijk}}{\sqrt{g}} \nabla_k
\bigl( R^j_i - \frac{1}{4} R \delta^j_i \bigr)
\end{eqnarray}
the Cotton tensor, and the covariant derivatives are defined with
respect to the spatial metric $g_{ij}$.
 $\epsilon^{ijk}$ is the totally
antisymmetric unit tensor, $\lambda$ is a dimensionless constant
and $\Lambda $ is a negative constant which is related to the
cosmological constant in the IR limit. Finally, the variables
$\kappa$, $w$ and $\mu$ are constants with mass dimensions $-1$,
$0$ and $1$, respectively.

Inserting a scalar field in the construction and imposing the
corresponding symmetries consistently, one results to the
following action
\cite{Calcagni:2009ar,Kiritsis:2009sh,Gao:2009ht}:
\begin{equation}
S_\phi = \int dtd^3x \sqrt{g} N \left[
\frac{3\lambda-1}{4}\frac{\dot\phi^2}{N^2}
+m_1m_2\phi\nabla^2\phi-\frac{1}{2}m_2^2\phi\nabla^4\phi +
\frac{1}{2}m_3^2\phi\nabla^6\phi -V_\phi(\phi)\right]~,
\end{equation}
where $V_\phi(\phi)$ acts as a potential term and $m_i$ are
constants (note that for simplicity we have absorbed the possible
term $-\frac{1}{2}m_1^2\phi^2$ inside $V_\phi(\phi)$). Clearly,
this is a simplified consideration which cannot cope with the
current knowledge of dark matter properties, but it allows for a
first investigation on the subject. Finally, we mention that one
could add the matter sector through a hydrodynamical approach,
adding a cosmological stress-energy tensor to the gravitational
field equations and demanding to recover the usual general
relativity formulation in the low-energy limit
\cite{Sotiriou:2009bx,Carloni:2009jc}. Since the scalar-field
approach to matter seems to have a better theoretical
justification, in this work we follow it. However, our results are
independent of the specific way that matter is incorporated, and
one could equivalently follow both the above matter-formulations.

 In principle one could include
additional scalars in the theory. The role of scalar fields in
cosmology has become crucial the last decades, mainly in inflation
\cite{Lindebook} or in dark energy phenomenology
\cite{Peebles:2002gy}, as well as in many other cases. However, in
the end of the day one should provide an explanation for their
appearance, and the usual approach is that the scalars arise from
some fundamental (probably higher-dimensional) theory of nature,
unknown up to now. Thus, although the additional scalar fields
have not a robust theoretical justification, nor it is clear how
they behave at the quantum level, it is still interesting to study
their effects on cosmology.

In this work we will allow for an additional scalar field, in
which we attribute the dark energy sector. Clearly, such an extra
field could put into question the renormalizability of the theory,
which must be examined in detail before the present model can be
considered as a realistic cosmology. However, the interesting
cosmological implications of the scenario at hand motivate us to
perform such a cosmological analysis, even with the
renormalizability subject, as well as the possible conceptual and
theoretical problems of Ho\v{r}ava-Lifshitz gravity itself
\cite{Charmousis:2009tc,Li:2009bg,Sotiriou:2009bx,Bogdanos:2009uj,Koyama:2009hc,Papazoglou:2009fj},
open for the moment. Thus, we add a second scalar $\sigma$, with
action
\begin{equation}
S_\sigma = \int dtd^3x \sqrt{g} N \left[
\frac{3\lambda-1}{4}\frac{\dot\sigma^2}{N^2}
+h_1h_2\sigma\nabla^2\sigma-\frac{1}{2}h_2^2\sigma\nabla^4\sigma +
\frac{1}{2}h_3^2\sigma\nabla^6\sigma -V_\sigma(\sigma) \right]~,
\end{equation}
where $V_\sigma(\sigma)$ accounts for the potential term of the
$\sigma$-field and $h_i$ are constants.

Now, in order to focus on cosmological frameworks, we impose an
FRW metric,
\begin{eqnarray}
N=1~,~~g_{ij}=a^2(t)\gamma_{ij}~,~~N^i=0,
\end{eqnarray}
with
\begin{eqnarray}
\gamma_{ij}dx^idx^j=\frac{dr^2}{1-kr^2}+r^2d\Omega_2^2~,
\end{eqnarray}
where $k=-1,0,1$ correspond to open, flat, and closed universe
respectively. In addition, we assume that the scalar fields are
homogenous, i.e $\phi\equiv\phi(t)$ and $\sigma\equiv\sigma(t)$.
By varying $N$ and $g_{ij}$, we obtain the equations of motion:
\begin{eqnarray}\label{Fr1}
H^2 &=&
\frac{\kappa^2}{6(3\la-1)}\left[\frac{3\la-1}{4}\,\dot\phi^2
+V_\phi(\phi)\right]+\nonumber\\
&+&\frac{\kappa^2}{6(3\la-1)}\left[\frac{3\la-1}{4}\,\dot\sigma^2
+V_\sigma(\sigma) -\frac{3\kappa^2\mu^2k^2}{8(3\lambda-1)a^4}
-\frac{3\kappa^2\mu^2\Lambda ^2}{8(3\lambda-1)}
 \right]+\nonumber\\
 &+&\frac{\kappa^4\mu^2\Lambda k}{8(3\lambda-1)^2a^2} \ ,
\end{eqnarray}
\begin{eqnarray}\label{Fr2}
\dot{H}+\frac{3}{2}H^2 &=&
-\frac{\kappa^2}{4(3\la-1)}\left[\frac{3\la-1}{4}\,\dot\phi^2
-V_\phi(\phi)\right]-\nonumber\\
&-&\frac{\kappa^2}{4(3\la-1)}\left[\frac{3\la-1}{4}\,\dot\sigma^2
-V_\sigma(\sigma) -\frac{\kappa^2\mu^2k^2}{8(3\lambda-1)a^4}
+\frac{3\kappa^2\mu^2\Lambda ^2}{8(3\lambda-1)}
 \right]+\nonumber\\
 &+&\frac{\kappa^4\mu^2\Lambda k}{16(3\lambda-1)^2a^2}\ ,
\end{eqnarray}
where we have defined the Hubble parameter as $H\equiv\frac{\dot
a}{a}$. Finally, the equations of motion for the scalar fields
read:
\begin{eqnarray}\label{phidott}
&&\ddot\phi+3H\dot\phi+\frac{2}{3\lambda-1}\frac{dV_\phi(\phi)}{d\phi}=0\\
\label{sdott}
&&\ddot\sigma+3H\dot\sigma+\frac{2}{3\lambda-1}\frac{dV_\sigma(\sigma)}{d\sigma}=0.
\end{eqnarray}

At this stage we can define the energy density and pressure for
the scalar fields. Concerning $\phi$, the corresponding relations
are
\begin{eqnarray}
&&\rho_\phi=\frac{3\la-1}{4}\,\dot\phi^2
+V_\phi(\phi)\equiv\rho_M\nonumber\\
&&p_\phi=\frac{3\la-1}{4}\,\dot\phi^2 -V_\phi(\phi)\equiv p_M,
\end{eqnarray}
and as we have mentioned they constitute the (dark) matter content
of the Ho\v{r}ava-Lifshitz universe. Concerning the dark energy
sector, we can define
\begin{equation}\label{rhoDE}
\rho_{DE}\equiv\frac{3\la-1}{4}\,\dot\sigma^2 +V_\sigma(\sigma)
-\frac{3\kappa^2\mu^2k^2}{8(3\lambda-1)a^4}
-\frac{3\kappa^2\mu^2\Lambda ^2}{8(3\lambda-1)}
\end{equation}
\begin{equation}
\label{pDE} p_{DE}\equiv\frac{3\la-1}{4}\,\dot\sigma^2
-V_\sigma(\sigma) -\frac{\kappa^2\mu^2k^2}{8(3\lambda-1)a^4}
+\frac{3\kappa^2\mu^2\Lambda ^2}{8(3\lambda-1)}.
\end{equation}
The first parts of these expressions, namely
$\frac{3\la-1}{4}\,\dot\sigma^2 +V_\sigma(\sigma)$ and
$\frac{3\la-1}{4}\,\dot\sigma^2 -V_\sigma(\sigma)$ correspond to
the energy density and pressure of the $\sigma$-field,
$\rho_\sigma$ and $p_\sigma$ respectively. The term proportional
to $a^{-4}$ is the usual ``dark radiation term'', present in
Ho\v{r}ava-Lifshitz cosmology
\cite{Calcagni:2009ar,Kiritsis:2009sh}. Note that it is present
even if we take the IR limit, that is it is a cosmological
artifact reflecting the novel features of Ho\v{r}ava-Lifshitz
gravitational background. Finally, the constant term is just the
explicit (negative) cosmological constant. Therefore, in
expressions (\ref{rhoDE}),(\ref{pDE}) we have defined the energy
density and pressure for the effective dark energy, which
incorporates the aforementioned contributions. We mention that we
could absorb the constant term inside the potential (equivalently
define an effective potential as
$\tilde{V}_\sigma(\sigma)=V_\sigma(\sigma)-\frac{3\kappa^2\mu^2\Lambda
^2}{8(3\lambda-1)}$), but we prefer to maintain it explicitly just
to keep truck of the origin of various terms, having in mind that
$V_\sigma(\sigma)$ must be sufficiently positive to assure for a
positive $\rho_{DE}$ as required by realistic cosmologies.

Using the above definitions, we can re-write the Friedmann
equations (\ref{Fr1}),(\ref{Fr2}) in the standard form:
\begin{eqnarray}
&&H^2 =
\frac{\kappa^2}{6(3\la-1)}\Big[\rho_M+\rho_{DE}\Big]+ \frac{\beta k}{a^2}\\
&&\dot{H}+\frac{3}{2}H^2 =
-\frac{\kappa^2}{4(3\la-1)}\Big[p_M+p_{DE}
 \Big]+ \frac{\beta k}{2a^2}.
\end{eqnarray}
In these relations we have defined
$\beta\equiv\frac{\kappa^4\mu^2\Lambda }{8(3\lambda-1)^2}$, which
is the coefficient of the curvature term. Additionally, we could
also define an effective Newton's constant and an effective light
speed \cite{Calcagni:2009ar,Kiritsis:2009sh}, but we prefer to
keep $\frac{\kappa^2}{6(3\la-1)}$ in the expressions, just to make
clear the origin of these terms in Ho\v{r}ava-Lifshitz cosmology.
Finally, note that using (\ref{phidott}),(\ref{sdott}) it is
straightforward to see that the aforementioned dark matter and
dark energy quantities verify the standard evolution equations:
\begin{eqnarray}\label{phidot2}
&&\dot{\rho}_M+3H(\rho_M+p_M)=0\\
\label{sdot2} &&\dot{\rho}_{DE}+3H(\rho_{DE}+p_{DE})=0.
\end{eqnarray}

\section{Cosmological implications and discussion}
\label{discuss}

In the previous section we formulated Ho\v{r}ava-Lifshitz dark
energy, that is we considered two scalar fields, one responsible
for dark matter fluid and one contributing to the dark energy
sector, in the framework of Ho\v{r}ava-Lifshitz  gravity. In this
section we examine the cosmological implications of this model and
in particular the dark energy phenomenology.

As usual, a central observable quantity is the dark energy
equation-of-state parameter, defined as:
\begin{equation}
\label{wDE}
w_{DE}\equiv\frac{p_{DE}}{\rho_{DE}}=\frac{\frac{3\la-1}{4}\,\dot\sigma^2
-V_\sigma(\sigma) -\frac{\kappa^2\mu^2k^2}{8(3\lambda-1)a^4}
+\frac{3\kappa^2\mu^2\Lambda
^2}{8(3\lambda-1)}}{\frac{3\la-1}{4}\,\dot\sigma^2
+V_\sigma(\sigma) -\frac{3\kappa^2\mu^2k^2}{8(3\lambda-1)a^4}
-\frac{3\kappa^2\mu^2\Lambda ^2}{8(3\lambda-1)}}.
\end{equation}
In the following we explore this expression in some cosmological
scenarios.

\subsection{Absent $\sigma$-field}

Let us first consider the simplified case of the complete absence
of the $\sigma$-field. We remind that even in this scenario, one
cannot avoid a constant potential term, in order to acquire a
positive-defined $\rho_{DE}$. In a sense, in the present
formulation of Ho\v{r}ava-Lifshitz dark energy, one cannot
eliminate the $\sigma$-field presence completely, since it will
always be (trivially but non neglectably) manifested itself
through a constant potential term. However, even if this case
seems special it leads to very interesting cosmological
implications, acting as a valuable example.

We start considering a flat ($k=0$) spacetime. In this case, as
expected, relation (\ref{wDE}) leads to $w_{DE}=-1$ that is we
obtain the simple cosmological constant universe. It is
interesting to see that the spacetime flatness sets to zero the
``dark radiation'' term that is present in (\ref{wDE}) (which is a
difference comparing to a similar term arising in braneworld
models \cite{Binetruy:1999ut}), and thus with the addition of the
disappearance of the $\sigma$-field terms we acquire the simple
result $w_{DE}=-1$.

As a next step we allow for a non-zero curvature. In this case the
dark radiation term appears, leading to a more complex
$w_{DE}$-behavior. In particular, setting $V_\sigma(\sigma)=V_0$
(the sufficiently positive constant discussed earlier), and
defining
\begin{equation}
\label{Votild}
 \tilde{V}_0=\frac{8V_0}{\kappa^2\mu^2}-\frac{3\Lambda^2}{3\lambda-1},
\end{equation}
(\ref{wDE}) leads to:
\begin{equation}
\label{wDE2} w_{DE}=\frac{-
\tilde{V}_0-\frac{1}{(3\lambda-1)a^4}}{
\tilde{V}_0-\frac{3}{(3\lambda-1)a^4}}.
\end{equation}
We mention that in principle, both $V_0$ and $\Lambda$ can be
arbitrary. However, since effectively $ \tilde{V}_0$ will play the
central role of dark energy in the current, observable universe,
one has to fine-tune $V_0$ and $\Lambda$ in order to
quantitatively lead to a very small $ \tilde{V}_0$, consistently
with observations.

A first observation is that the ``running'' behavior of
Ho\v{r}ava-Lifshitz background is reflected in the ``running''
behavior of $w_{DE}$. Furthermore, surprisingly enough we observe
that $w_{DE}<-1$ always, that is we result to an effective dark
energy presenting a phantom behavior. This behavior is a pure
effect of the dark radiation term, and enlightens the discussion
about the novel implications of Ho\v{r}ava-Lifshitz dark energy.
Progressively, as the scale factor increases, dark radiation
dilutes and $w_{DE}$ asymptotically goes to $-1$ at very large
times. Thus, although presenting a phantom behavior, this scenario
is free of a Big-Rip
\cite{Caldwell03,Gonzalezl03,Gonzalezl03b,Nojiri:2005sx}.

The cosmological evolution is also very interesting in the other
``direction'', that is going to small scale factors. Due to the
presence of the dark radiation term, one can easily see from the
Friedmann equations (\ref{Fr1}), (\ref{Fr2}) that at some
particular moment $\dot{H}$ changes sign leading to a bounce
\cite{Brandenberger:2009yt,Brandenberger:2009ic}. Therefore,
$\rho_{DE}$ will never become negative and the universe will never
become singular.

\subsection{Flat universe with $\sigma$-field}

Let us now consider a flat universe, with the presence of both
matter and $\sigma$ fields. In this case we obtain:
\begin{equation}
w_{DE}=\frac{\frac{3\la-1}{4}\,\dot\sigma^2 -V_\sigma(\sigma)
+\frac{3\kappa^2\mu^2\Lambda
^2}{8(3\lambda-1)}}{\frac{3\la-1}{4}\,\dot\sigma^2
+V_\sigma(\sigma) -\frac{3\kappa^2\mu^2\Lambda
^2}{8(3\lambda-1)}}.
\end{equation}
As we see, $w_{DE}$ presents the ``running'' behavior that is
inherited from the Ho\v{r}ava-Lifshitz background. Clearly, in the
IR fixed point $\la=1$, and under the absorption of the constant
term in the potential, we re-obtain the usual equation-of-state
parameter of standard dark energy formalism
\cite{quintess,quintess2,quintess3}. Therefore, since dark energy
observations are by far inside the IR, Ho\v{r}ava-Lifshitz
artifacts of this scenario would not be observable and the model
is not distinguishable from standard dark energy paradigms.
Finally, concerning the cosmological behavior of $w_{DE}$, we see
that it lies always at the quintessence regime ($-1<w_{DE}$) and
the evolution is free of a Big Rip. The reason for these features
is that the universe flatness kills the dark radiation term and
thus it disappears the novel and unexpected phenomena.

\subsection{Non-flat universe with $\sigma$-field}

In this scenario the dark energy equation-of-state parameter is
given by (\ref{wDE}) in full generality. Similarly to the previous
subsection, we observe that $w_{DE}$ depends explicitly on
$\lambda$, with its value at $\la=1$ corresponding to the IR
limit. However, the Ho\v{r}ava-Lifshitz nature of the model is
manifested in an unexpected cosmological implication which is
independent of the energy scale. Namely, the effective dark energy
can experience the crossing of the phantom divide $-1$, for a
suitable potential, as can be easily observed from (\ref{wDE}).
Amazingly, this is achieved despite taking the IR limit of the
theory. Thus, in this case, artifacts of Ho\v{r}ava-Lifshitz
gravity could be detected through dark energy observations.
However, one still cannot distinguish between this model and
alternative models that allow for the realization of $w_{DE}<-1$
phase, such are modified gravity
\cite{ordishov,ordishov2,Nojiri:2003ft,Elizalde:2004mq,Nojiri:2003ni}
or models with phantom \cite{Caldwell03,phant,phant2,phan5} or
quintom fields
\cite{quintom,quintom2,quintom3,quintom4a,quintom4}.

 In order to provide an explicit but general example of the
aforementioned crossing behavior we consider power-law solutions
of the form $\sigma(t)=c_\sigma t^q$ and $a=c_a t^p$, which
satisfy the Friedmann and field equations for appropriately
constructed potentials. In this case, the crossing is realized at
\begin{equation}
a_{cr}=c_a\left[\frac{\kappa\mu}{c_\sigma c_a^2
q(3\la-1)}\right]^{\frac{p}{q+2p-1}}.
\end{equation}
We mention that according to the parameter values, the
$-1$-crossing can be realized at an arbitrary scale factor, either
at early or at late times.

\subsection{Bounce, turnaround and cyclic behavior}

Apart from interesting dark energy phenomenology, the general
scenario of the presence of both scalar fields in a non-flat
universe allows for interesting cosmological behavior at all
epochs. In particular, at small scale factors the role the dark
radiation term inside Ho\v{r}ava-Lifshitz dark energy is enhanced
and one naturally obtains a bounce, thus avoiding the possible
singular behavior of the universe. We mention that due to the
presence of the $\sigma$-field the bounce is always obtained and
it is not restricted to slowly-diluted matter contents
\cite{Brandenberger:2009yt,Brandenberger:2009ic}. On the other
hand, one can easily construct a large sub-class of
$\sigma$-potentials which at late times allow the negative
cosmological constant to dominate the evolution (the positivity of
$\rho_{DE}$ is not affected) and thus to trigger a turnaround. In
other words, the competition of the dark radiation and
cosmological constant terms, can make Ho\v{r}ava-Lifshitz dark
energy naturally addressing cyclic cosmology. A construction of
such a model is left for future investigation.

\section{Conclusions}
\label{concl}

In this work, using an additional canonical scalar field, we have
formulated Ho\v{r}ava-Lifshitz cosmology with an effective dark
energy sector. Ho\v{r}ava-Lifshitz dark energy, due to the
inherited features from the novel Ho\v{r}ava-Lifshitz
gravitational background, proves to present very interesting
behaviors, possessing a varying equation-of-state parameter,
exhibiting phantom behavior and allowing for a realization of the
phantom divide crossing. Although according to current
observations one cannot distinguish it from well-studied scenarios
such as quintessence, that is from General Relativity with an
extra scalar field, its novelty is that it possesses an improved
gravitational background with UV completeness. Moreover,
Ho\v{r}ava-Lifshitz dark energy can always guarantee for a bounce
at small scale factors, offering a simple way to avoid singular
cosmological evolution. Furthermore, it can easily accept simple
solution sub-classes where it can trigger both bounce and
turnaround and thus lead naturally to cyclic cosmology. These
features were the motivations of the present work.

The existence of a second scalar field offers larger possibilities
to control the aforementioned behaviors, either at small or large
scales factors, or even simultaneously for both cosmological
regimes. Furthermore, such a construction could alleviate the
undesirable phenomena that could arise in the gravitational sector
\cite{Charmousis:2009tc,Li:2009bg}. Although for simplicity we
have performed the analysis under the detailed-balance condition,
the basic features of the construction are independent of that and
are expected to be present even if we relax this constraint. On
the contrary, a similar investigation without detailed-balance
will make the construction more robust and free of possible
problems that detailed-balance could bring in Ho\v{r}ava-Lifshitz
gravity itself \cite{Charmousis:2009tc,Kiritsis:2009sh}. Lastly,
note that the presence of the Cotton tensor prevents high-momentum
pathological behavior \cite{Kiritsis:2009sh}. However, it would be
interesting to use Renormalization Group methods to study the
running of the quantities, leading to improved dynamics, similarly
to the earlier works \cite{Bonanno:2001xi,Bonanno:2001hi}. But
such an investigation lies beyond the scope of the present work.

Finally, we mention that the present work is just a first approach
on the subject of dark energy in Ho\v{r}ava-Lifshitz cosmology,
definitely far from a solution to the dark energy problem. It is
interesting to note that one faces the same two ways to acquire
acceleration as in ``conventional'' cosmology. In particular, in
the later it is widely known that one can either use an additional
scalar field (quintessence \cite{quintess,quintess2,quintess3} or
phantom one \cite{Caldwell03,phant}), or modify the gravitational
sector itself
\cite{ordishov,ordishov2,Nojiri:2003ft,Elizalde:2004mq,Nojiri:2003ni}.
Similarly, in the case of Ho\v{r}ava-Lifshitz cosmology, one can
either use an additional scalar field, an approach followed in
this work, or he can modify and generalize the gravitational
action of Ho\v{r}ava-Lifshitz gravity
\cite{Nojiri:2009th,Kluson:2009rk,Chaichian:2010yi}. Each approach
exhibts its own advantages and disadvantages.

\begin{acknowledgments}
The author wishes to thank Yi-Fu Cai for crucial discussions,
Elias Kiritsis and Gianluca Calcagni for valuable comments, an
anonymous referee for useful advices, and Institut de Physique
Th\'eorique, CEA, for the hospitality during the preparation of
the present work.
\end{acknowledgments}

\end{document}